%% file: arxiv.tex
\documentclass[10pt,letterpaper]{article}

\input{Paper_Style.sty}

\input{Alphabet_Commands.tex}

\input{Math_Commands.tex}

\input{Paper_Specific_Commands.tex}
\usepackage{adjustbox}
\usepackage{pdflscape}
\usepackage{caption}
\usepackage{subcaption}
\usepackage{colortbl}
\usepackage[table]{xcolor}
\definecolor{lightgray}{gray}{0.95}
\usepackage[top=0.85in,footskip=0.75in]{geometry} 

\usepackage{amsmath,amssymb}

\usepackage{comment}

\usepackage{changepage}

\usepackage{textcomp,marvosym}

\usepackage{cite}

\usepackage{nameref,hyperref}

\usepackage[right]{lineno}

\usepackage{microtype}
\DisableLigatures[f]{encoding = *, family = * }

\usepackage[table]{xcolor}

\usepackage{array}

\newcolumntype{+}{!{\vrule width 2pt}}

\newlength\savedwidth

\raggedright
\usepackage[aboveskip=1pt,labelfont=bf,labelsep=period,justification=raggedright,singlelinecheck=off]{caption}

\makeatletter
\renewcommand{\@biblabel}[1]{\quad#1.}
\makeatother

\usepackage{lastpage,fancyhdr,graphicx}
\usepackage{epstopdf}
\fancyheadoffset[L]{0.75in}
\fancyfootoffset[L]{0.75in}
\fancyheadoffset[R]{0.75in}
\fancyfootoffset[R]{0.75in}
\begin{document}

\vspace*{0.2in}

\begin{flushleft}
{\Large
\textbf\newline{Reactive means in the Iterated Prisoner's Dilemma} 
}
\newline
\\
Grant Molnar\textsuperscript{1},
Caroline Hammond\textsuperscript{1},
Feng Fu\textsuperscript{1,2} 
\\
\bigskip
\textbf{1} Department of Mathematics, Dartmouth College, Hanover, NH 03755, USA\\

\textbf{2} Department of Biomedical Data Science, Geisel School of Medicine at Dartmouth, Lebanon, NH 03756, USA
\bigskip
\end{flushleft}

\section*{Abstract}
    The Iterated Prisoner's Dilemma (IPD) is a well studied framework for understanding direct reciprocity and cooperation in pairwise encounters. However, measuring the morality of various IPD strategies is still largely lacking. Here, we partially address this issue by proposing a suit of plausible morality metrics to quantify four aspects of justice. We focus our closed-form calculation on the class of reactive strategies because of their mathematical tractability and expressive power. 
    We define reactive means as a tool for studying how actors in the IPD and Iterated Snowdrift Game (ISG) behave under typical circumstances. We compute reactive means for four functions intended to capture human intuitions about ``goodness'' and ``fair play''. Two of these functions are strongly anticorrelated with success in the IPD and ISG, and the other two are weakly anticorrelated with success. Our results will aid in evaluating and comparing powerful IPD strategies based on machine learning algorithms, using simple and intuitive morality metrics.

\section{Introduction}
\label{Section: Introduction}
Iterated $2 \times 2$ games, most notably the Iterated Prisoner's Dilemma (IPD), have been the objects of intensive study at least since Axelrod's classical experiments \cite{Axelrod}. Much research has been devoted to determining which strategies perform well under various circumstances \cite{Nowak, press2012iterated} \cite{Singer}; the IPD has been studied in the presence of noise \cite{Nowak}, social dynamics \cite{Prisoner}, and with other variations~\cite{Axel2, Prisoner,bazzan2002evolution,hirshleifer1989cooperation,wu1995cope}. The framework of IPD and evolutionary game theory more generally have offered profound insights into understanding the evolution of cooperation~\cite{axelrod1981evolution,nowak2006five,grujic2010social,perc2017statistical,ezaki2016reinforcement,wang2022decentralized}.  

In particular, the discovery of zero-determinant (ZD) strategies by Press and Dyson has greatly reinvigorated the field with brand new perspectives~\cite{press2012iterated,stewart2013extortion,chen2022intricate,chen2022outlearning,ichinose2018zero}. ZD strategies are able to unilaterally enforce a linear relationship between their own average payoff and that of their co-player. An extortionate ZD player can thus take advantage of deliberately prescribed ZD strategies to demand an unfair share from their mutual interactions. Motivated by this fact, researchers have attempted to classify IPD strategies, for example, into partners versus rivals, by their capacity of fostering mutual cooperation or securing unilateral winning~\cite{akin2015you,akin2016iterated,hilbe2018partners}. This dichotomic classification has a natural extension to the idea of morality. Human behavior is not solely guided by the desire to win, but also by moral values and judgments~\cite{tabibnia2007fairness}. While such a classification of ZD strategies might be enlightening, there is also a strong need for studying the morality of IPD strategies more broadly.  

In \cite{Singer}, Singer-Clark investigates the question of which IPD strategies are the ``most moral" using a different methodology. Under this framework, a player in an IPD treats their competitor well if they cooperate a large proportion of the time. Singer-Clark uses eigenvalues on a population of such strategies to define two measures, EigenJesus and EigenMoses, for which strategy was ``most moral''. This is a fascinating approach, but it has some serious drawbacks. One is that it does not incorporate noise or error, making it less applicable to real-world scenarios. It is also unclear how to generalize Singer-Clark's methodology from $2 \times 2$ games whose choices have a clear social valence, like the Prisoner's Dilemma and Snowdrift, to other more complicated or nuanced games. A third drawback is the atemporality of these metrics: they do not pay attention to which player defected first, only which one defected more.  
Most seriously from our perspective, Singer-Clark's eigenvalue-based  morality is socially contingent. That is, for Singer-Clark, an actor's morality depends on who they are playing against. 
It is natural to ask: is there some way of determining how a player behaves without relying on these variable social contingencies?  We pursue this line of inquiry by introducing another set of metrics to assess the morality of IPD and Iterated Snowdrift Game (ISG) strategies.

For the rest of Section \ref{Section: Introduction}, we will introduce the mechanics of the games and strategies analyzed in this paper along with defining the reactive mean. An analysis of our results is included. Section \ref{Section: Metrics of Justice} defines the player-oriented functions that will be analyzed. Section \ref{Section: Results and Analysis} provides some statistics about the reactive means for our functions of interest. The final section, Section \ref{Section: Future Work}, discusses further applications of reactive means. The explicit calculations for the reactive means are provided in the \nameref{Section: Appendix}.

\subsection{Games of Interest}
The Prisoner's Dilemma is a simple $2 \times 2$ game. Fix a tuple of real numbers $\vec{R} = (R_{CC}, R_{CD}, R_{DC}, R_{DD})$ such that 
\[
R_{DC} > R_{CC} > R_{DD} > R_{CD} \ \text{and}  \ 2R_{CC} > R_{CD} + R_{DC}.
\]
Each player chooses to cooperate ($C$) or defect ($D$). If both players cooperate, they each receive a reward $R_{CC}$. If one player cooperates and the other player defects, then the cooperating player receives reward $R_{DC}$, and the defecting player receives reward $R_{CD}$. If both players defect, they each receive reward $R_{DD}$. By construction, players are collectively best off when they both cooperate, but are individually better off when they individually defect. Axelrod chose $\vec{R} = (3,0,5,1)$ for his famous tournaments \cite{Axelrod}, and so these values are standard in much of the literature, but any $2 \times 2$ game with payoffs satisfying the inequalities above qualifies as a Prisoner's Dilemma. An IPD is simply a Prisoner's Dilemma played repeatedly between the same two individuals. 

The Snowdrift Game is formally almost identical to the Prisoner's Dilemma but with a different payoff structure~\cite{van2015importance}. As above, we fix a tuple of real numbers $\vec{R} = (R_{CC}, R_{CD}, R_{DC}, R_{DD})$, and each player can choose to cooperate ($C$) or defect ($D$), with commensurate payoffs. However, we now ask that
\[
R_{DC} > R_{CC} > R_{CD} > R_{DD} \ \text{and} \ 2R_{CC} = R_{CD} + R_{DC}.
\]
Therefore, no value is destroyed when a player defects against a cooperative adversary, and a player is better of cooperating than defecting against a defector. In accordance with \cite{van2015importance} and \cite{Axelrod}, it is common to use $\vec{R} = (3,1,5,0)$ for the Snowdrift game. The following scenario provides one interpretation of this game: two individuals are driving up an icy road when they discover a snowdrift cutting the avenue off. As long as at least one of them digs, the snowdrift will be removed and both can keep driving. However, neither individual enjoys the process of digging. The game of ``Chicken" provides another interpretation of the Snowdrift Game. From this perspective, the players are car drivers heading towards each other to prove their courage. If either one pulls out early, then both survive, and the player who didn't flinch also accrues accolades and honor. If both pull out simultaneously, the honor is split evenly between them. If neither pulls out, they both die. Comparably to an IPD, an ISG is a Snowdrift Game played repeatedly between the same two opponents.

\subsection{Strategies of Interest}
To investigate this idea, we analyze the behavior of a specific category of strategies. A \textbf{reactive memory one} strategy $A$ is a triple $A = \parent{\pr 0, \pr C, \pr D} \in [0,1]^3$, where $\pr 0$ is the probability that $A$ cooperates on the first round of the game, and $\pr x$ is the probability that $A$ cooperates if their opponent made move $x$ in the previous round.

Let $A = \parent{\pr 0, \pr C, \pr D }$ and $A\prm = \parent{\prprm 0, {\prprm C}, {\prprm C}}$ be reactive memory one strategies. We define $\pr C (A, A\prm) \coloneqq \pr C$, and we define $\pr D(A, A\prm) \coloneqq \pr D$. 

If $f(A,A\prm)$ is a function depending on $A$ and $A\prm$, we write $f\prm(A, A\prm) \coloneqq f(A\prm, A)$ for the function that interchanges the roles of $A$ and $A\prm$. For instance, $\pr C(A, A\prm) = \pr C$, and $\pr C\prm(A, A\prm) = \pr C(A\prm, A) = \pr C \prm$. 

We write 
\[
	\vec{\pi}_N = \vec{\pi}_N( A , A\prm ) 
	= \begin{pmatrix}
	{\statpr {CC}{,N}} & {\statpr {CD}{,N}} & {\statpr {DC}{,N}} & {\statpr {DD}{,N}}
	\end{pmatrix}
\]
for the probability distribution of cooperation and defection for $A$ and $A\prm$ after $N$ rounds, and observe
\[
\vec{\pi}_0 = \vec{\pi}_0(A, A\prm) = \begin{pmatrix}
\pr 0 \prprm 0 & \pr 0 (1-\prprm 0) & (1-\pr 0) \prprm 0 & (1-\pr 0)(1 - \prprm 0)
\end{pmatrix}.
\]
We also write 
\[
	P(A, A\prm) = \begin{pmatrix}
	\pr C \prprm C & \pr C (1 - \prprm C) & (1 - \pr C) \prprm C & (1 - \pr C) (1 - \prprm C) \\
	\pr D \prprm C & \pr D (1 - \prprm C) & (1 - \pr D) \prprm C & (1 - \pr D) (1 - \prprm C) \\
	\pr C \prprm D & \pr C (1 - \prprm D) & (1 - \pr C) \prprm D & (1 - \pr C) (1 - \prprm D) \\
	\pr D \prprm D & \pr D (1 - \prprm D) & (1 - \pr D) \prprm D & (1 - \pr D) (1 - \prprm D)
	\end{pmatrix}
\]
for the transition matrix of the Markov process indicated above. Clearly, we have $\vec{\pi}_N = \vec{\pi}_0 P^N$ for all $N \geq 0$. If $P$ is mixing, then there is a unique steady-state distribution 
\[
\vec{\pi}_\infty = \vec{\pi}_\infty(A, A\prm) = \begin{pmatrix}
\statpr CC & \statpr CD & \statpr DC & \statpr DD
\end{pmatrix}
\]
for $A$ and $A\prm$. Let $\coop = \coop (A, A\prm)$ denote the long-run probability that $A$ cooperates in any given round; thus, $\coopprm$ is the probability that $A\prm$ cooperates in any given round. In \cite{Nowak}, Nowak proved that
\[
	\pi_\infty = \begin{pmatrix}
\coop \coopprm & \coop (1 - \coopprm) & (1 - \coop) \coopprm & (1 - \coop ) (1 - \coopprm)
\end{pmatrix},
\]
and gave the following formulas for $\coop$ and $\coopprm$:

\begin{align}
\coop &= \frac{\prprm D \response + \pr D}{1 - \response \responseprm} \label{c defined} \\
\coopprm &= \frac{\pr D \responseprm + \prprm D}{1 - \response \responseprm}. \label{cprm defined}
\end{align}

Here $\response \coloneqq \pr C - \pr D$ and $\responseprm \coloneqq \prprm C - \prprm D$. The quantity $\response$ measures the \textbf{responsiveness} of $A$; that is, the degree to which $A$ treats adversaries who cooperate better than adversaries who defects. Writing $\coop = \coop (A, A\prm)$ as a function of $A$ and $A\prm$, we find $\coopprm(A, A\prm) = \coop(A\prm, A)$, as we should expect.

Let $f$ be an integrable function of two strategies $A$ and $A\prm$, and write $X = [0,1]^3$. We define 
\[
	\Mean f(A) \coloneqq \frac{1}{\vol(X)} \int\limits_X f(A, Q) \dif Q
\] 
to be the \textbf{reactive mean of $f$ for $A$}. Note that $\vol(X) = 1$, so in fact
\begin{align*}
	\Mean f(A) &= \int\limits_X f(A, Q) \dif Q
\end{align*}
The quantity $\Mean f(A)$ measures the expected value of $f$ when an adversary for $A$ is chosen uniformly at random from the set of reactive memory one strategies. If $f$ is independent of $\pr 0$, then 
\begin{align}
	\Mean f(A) = \int\limits_0^1 \int\limits_0^1 f(\pr C, \pr D, \prprm C, \prprm D) \dif {\prprm C} \dif {\prprm D}. 
 \label{fA defined independent of p0}
\end{align}

\subsection{Summary of Results}
We were able to show that a player's score in both the IPD and ISG is negatively correlated with all four of the metrics for justice delineated in Section \ref{Section: Metrics of Justice} to different extents. Our methods are especially exciting because they give objective measures of the behavior of actors. Unlike the EigenMoses and EigenJesus metrics calculated in \cite{Singer}, the reactive means of asymptotic niceness, long-term cooperation rate, responsiveness, and reciprocity (see Section \ref{Section: Metrics of Justice} below) do not depend on the behaviors of individual opponents. As a result, players can be assigned a strict level of morality that does not change with the population of opponents. In addition, the measures of morality obtained in this paper apply to opponents of every possible reactive memory one strategy, of which there are infinitely many. As long as the values of $\pr C$ and $\pr D$ are known, it is straightforward to evaluate any of these morality functions. Also, the methodology used in this paper incorporates noise to some degree since a memory-one strategy with added noise is essentially just another memory one strategy. Of course, this does not incorporate all forms of noise since it can cause the strategy to shift dynamically, but it does make the results more realistic. 

\section{Model and Methods}
\label{Section: Metrics of Justice}

\paragraph{Metrics of Justice.}
In modern parlance, ``justice'' means ``retribution''. Historically, however, justice was equated with social morality writ large, subsuming concepts like fair play, and treating other people well. We take inspiration from these intuitions to enumerate a few loose criteria for just actions.

\begin{enumerate}
	\item A player is just insofar as they treat kind players well. \label{Justice as reward}
	\item A player is just insofar as they treat other players well. \label{Justice as kindness}
	\item A player is just insofar as they treat other players well when the others treat them well. \label{Justice as positive reciprocity}
	\item A player is just insofar as they treat other players as the others treat them. \label{Justice as reciprocity}
\end{enumerate}
Inspired by these intuitions, we develop various metrics which correspond to our folk sense of justice.

\subsection{Asymptotic Niceness}
Axelrod observed that the most successful strategies in his tournament were ``nice'' in the sense that they did not defect before their adversaries did \cite[p.10]{Axelrod}. This is not an especially useful notion for us, however, because it depends intimately on $p_0$, whereas our focus is on long-run behavior. Thus we define the \textbf{(asymptotic) niceness} $\nice(A, A\prm)$ of $A$ against $A\prm$ to be the long-run probability that if $A$ and $A\prm$ cooperate in the same round, $A\prm$ subsequently defects before $A$. Thus if $A$ is a reactive memory one strategy, and $\nice(A, A\prm) = 1$ for every reactive memory one strategy $A\prm$, then $\pr C(A) = 1$, and if $\nice(A, A\prm) = 0$ for every reactive memory one strategy $A\prm$, then $\pr C(A) = 0$.

\subsection{Reciprocity}
Let $T \coloneqq (1, 1, 0)$ denote the strategy that begins by cooperating and then reciprocates the last move of their adversary; this famous strategy is referred to as ``Tit-for-Tat''. Axelrod observed that $T$ fared better than any other strategy in his tournaments. Extensive research has gone into when and how $T$ succeeds against other strategies \cite{Axelrod, Singer}, but $T$ also perfectly exemplifies a willingness to reciprocate the actions of its adversaries. In a sense, $T$ acts with perfect justice. Consider by contrast ``the Bully'' $B \coloneqq (0, 0, 1)$, which begins by defecting and then defects against cooperation and cooperates against defection. The Bully exploits those who are willing to cooperate with it, while submitting to and cooperating with those who defect against it. In a sense, $B$ is the opposite of $T$; indeed, the coordinates $(0, 0, 1)$ are maximally distant from the coordinates $(1, 1, 0)$ in the unit square. Moreover, $B$ behaves in a way that intuitively parses as ``evil'': preying on the kind, and capitulating to the cruel. We define the \textbf{reciprocity} $\tft(A, A\prm)$ of $A$ with $A\prm$ to be the long-run probability that $A$'s move is the same as $A\prm$'s previous move. Thus if $A$ is a reactive memory one strategy, and $\tft(A, A\prm) = 1$ for every reactive memory one strategy $A\prm$, then $\pr C(A) = 1$ and $\pr D (A) = 0$, and if $\tft(A, A\prm) = 0$ for every reactive memory one strategy $A\prm$, then $\pr C(A) = 0$ and $\pr D (A) = 1$.

\subsection{Functions of Interest}
Define $\scorepr(A, A\prm)$ as the long-run average score that $A$ earns each round against $A\prm$ in the IPD. For instance, if $T$ is Tit-for-Tat, and $B$ is the Bully, then $\scorepr(T, T) = 3$, and $\scorepr(B, B) = 2$ using the values from classic literature mentioned in Section \ref{Section: Introduction}. Likewise, let $\scoresn(A, A\prm)$ denote the long-run average score that $A$ earns each round against $A\prm$ in the ISG. For instance, if $T$ is Tit-for-Tat, and $B$ is the Bully, then $\scoresn(T, T) = 3$, and $\scoresn(B, B) = 3/2$.

Let 
\[
	\calF \coloneqq \set{\pr C, \pr D, \nice, \coop, \response, \tft, \scorepr, \scoresn, \statpr CC, \statpr CD, \statpr DC, \statpr DD}.
\]
The set $\calF$ comprises our functions of interest for this paper. The functions $\pr C,$ $\pr D,$ $\statpr CC,$ $\statpr CD,$ $\statpr DC,$ $\statpr DD$ comprise the fundamental building blocks for the behavior of strategy $A$ against strategy $A\prm$. The functions $\nice$ and $\coop$ measure the degree to which $A$ is kind to $A\prm$, in the sense of not initiating defection (for $\nice$), or cooperating (for $\coop$). These functions reflect the metrics of justice defined in \ref{Justice as reward} and \ref{Justice as kindness}, respectively. The functions $\response$ and $\tft$ measure the degree to which $A$ reciprocates the actions of $A\prm$; in other words, the degree to which $A$ asymptotically follows the Tit-for-Tat strategy. These functions reflect the ideas of \ref{Justice as positive reciprocity} and \ref{Justice as reciprocity}. The functions $\scorepr$ and $\scoresn$ measure the success of $A$ against $A\prm$ in the IPD and ISG, respectively.

These quantities are intimately interconnected, and can each be expressed in terms of $\pr C$ and $\pr D$. For ease of notation, we suppress dependence on $A$ and $A\prm$ in the equations below.

\begin{align*}
    \nice &= \frac{\pr C \parent{1 - \prprm C}}{1 - {\pr C} {\prprm C}} = 1 - \frac{1 - \pr C}{1 - \pr C \prprm C}, \\
	\tft &= \pr C \cdot \coopprm + (1 - \pr D) \cdot (1 - \coopprm), \\
    \score &= R_{CC} \cdot \statpr CC + R_{CD} \cdot \statpr CD + R_{DC} \cdot \statpr DC + R_{DD} \cdot \statpr DD.
\end{align*}

For each function of interest $f$, we have an explicit formula for $\Mean f$; however, these formulas are generally ungainly and unedifying, so we have relegated them to the \nameref{Section: Appendix}, where they are used to produce cleaner data.

\section{Results and Analysis}
\label{Section: Results and Analysis}
\subsection{Heat Maps}
For each function of interest $f \in \calF$, we have a heat map for $\Mean f$ pictured below. $\Mean f$ is graphed with white (for low values) and dark purple (for high values) as a function of $\pr C$ and $\pr D$. The scale for each heat map is given to its right. Heat maps of the complements $\Mean{\niceprm}$, $\Mean{\coopprm}$, $\Mean{\tftprm}$, $\Mean{\scoreprm_{\text{pr}}}$, and $\Mean{\scoreprm_{\text{sn}}}$ are also included. Neither heat maps nor statistical analyses of $\Mean{\prprm C}$, $\Mean{\prprm D}$, and $\Mean{r \prm}$ are included because these are constant values. Additionally, $\Mean{\statprprm CC}$, $\Mean{\statprprm CD}$, $\Mean{\statprprm DC}$, and $\Mean{\statprprm DD}$ are not included since $\statprprm CC$=$\statpr CC$, $\statprprm DD$=$\statpr DD$, $\statprprm CD$=$\statpr DC$, and $\statprprm DC$=$\statpr CD$, making additional analyses of these complements redundant. In the heat maps for $\Mean{\score_{pr}}$ and $\Mean{\score_{sn}}$, the different values of $R$ correspond to those used in classic literature referred to in Section \ref{Section: Introduction}.

\begin{figure}
     \centering
     \begin{subfigure}[b]{0.49\textwidth}
         \centering
         \includegraphics[width=\textwidth]{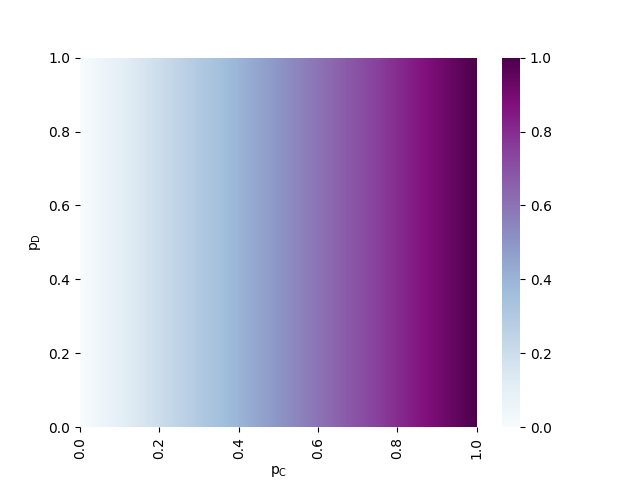}
         \caption{$\Mean{\pr C}$}
         \label{fig:pc}
     \end{subfigure}
     \hfill
     \begin{subfigure}[b]{0.49\textwidth}
         \centering
         \includegraphics[width=\textwidth]{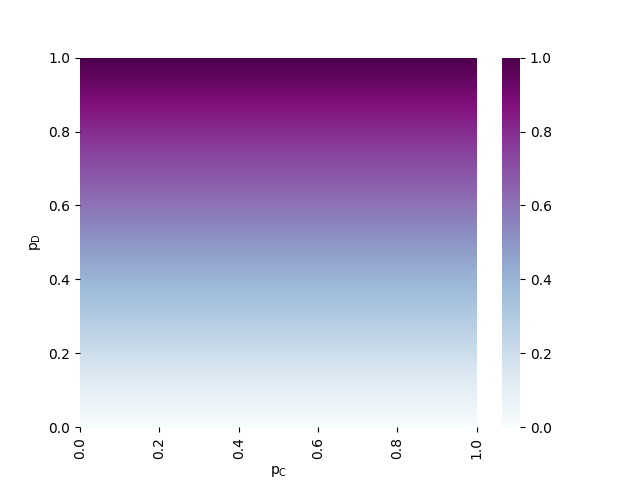}
         \caption{$\Mean{\pr D}$}
         \label{fig:pd}
     \end{subfigure}
     \hfill
     \begin{subfigure}[b]{0.51\textwidth}
         \centering
         \includegraphics[width=\textwidth]{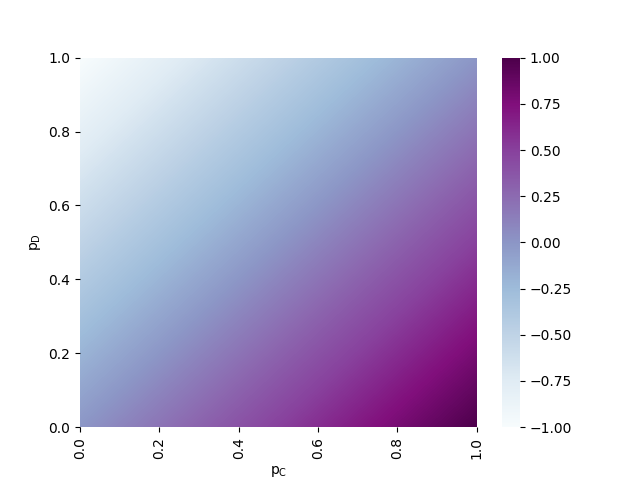}
         \caption{$\Mean{r}$}
         \label{fig:r}
     \end{subfigure}
     \hfill
     \begin{subfigure}[b]{0.49\textwidth}
         \centering
         \includegraphics[width=\textwidth]{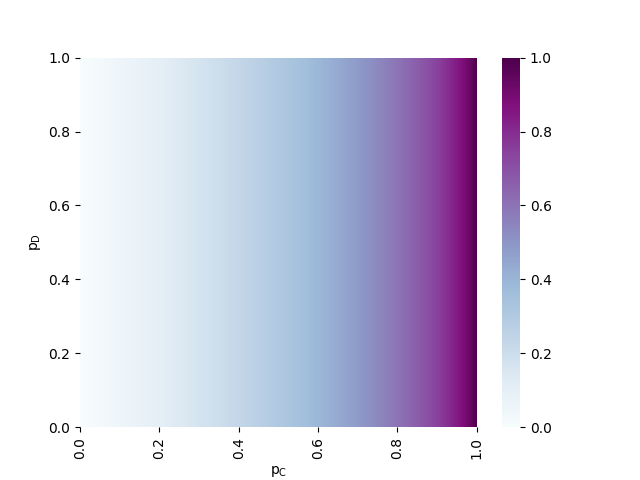}
         \caption{$\Mean{\nice}$}
         \label{fig:n}
     \end{subfigure}
     \hfill
     \begin{subfigure}[b]{0.49\textwidth}
         \centering
         \includegraphics[width=\textwidth]{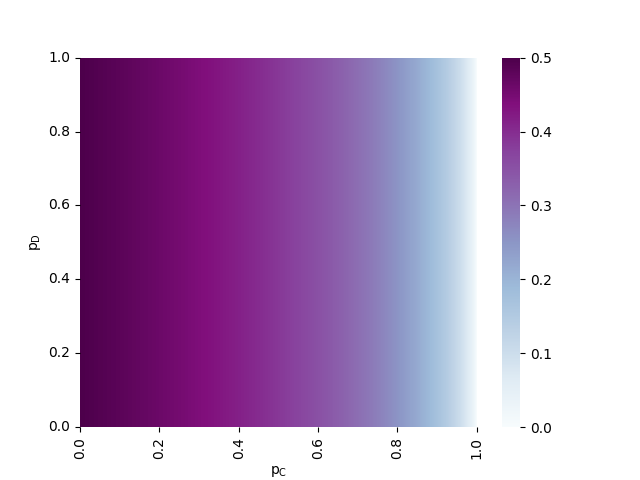}
         \caption{$\Mean{\niceprm}$}
         \label{fig:n'}
     \end{subfigure}
     \\
     \hfill
        \caption{Heat diagrams for $\Mean{\pr C}$, $\Mean{\pr D}$, $\Mean{r}$, $\Mean{\nice}$, and $\Mean{\niceprm}$ (a,b,c,d,e) with respect to $\pr C$ and $\pr D$}
        \label{fig:three graphs}
\end{figure}

\begin{figure}
     \centering
     \begin{subfigure}[b]{0.49\textwidth}
         \centering
         \includegraphics[width=\textwidth]{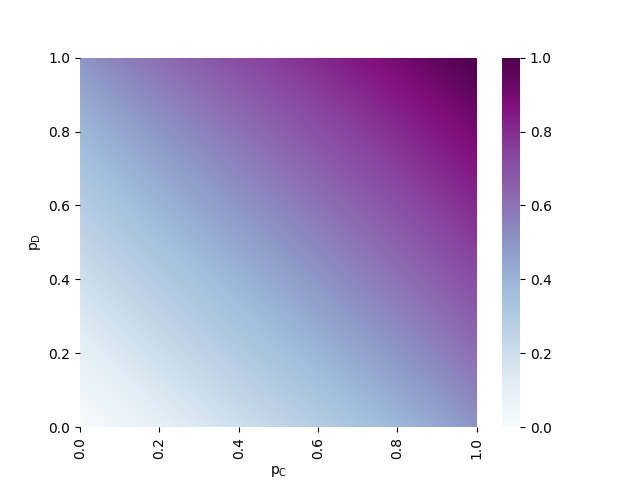}
         \caption{$\Mean{c}$}
         \label{fig:cc}
     \end{subfigure}
     \hfill
     \begin{subfigure}[b]{0.49\textwidth}
         \centering
         \includegraphics[width=\textwidth]{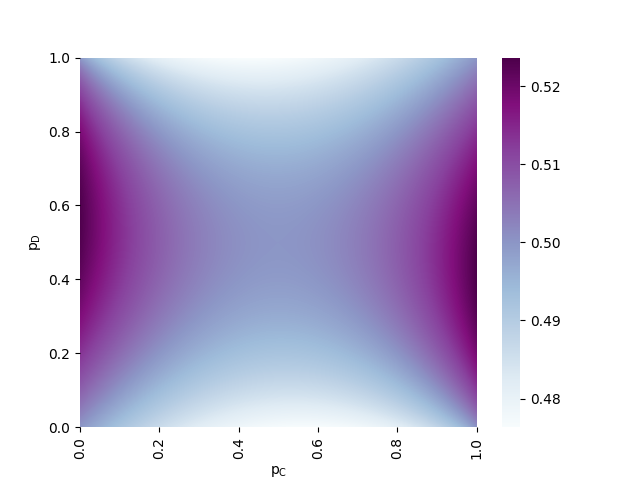}
         \caption{$\Mean{c \prm}$}
         \label{fig:c'}
     \end{subfigure}
     \hfill
     \begin{subfigure}[b]{0.49\textwidth}
         \centering
         \includegraphics[width=\textwidth]{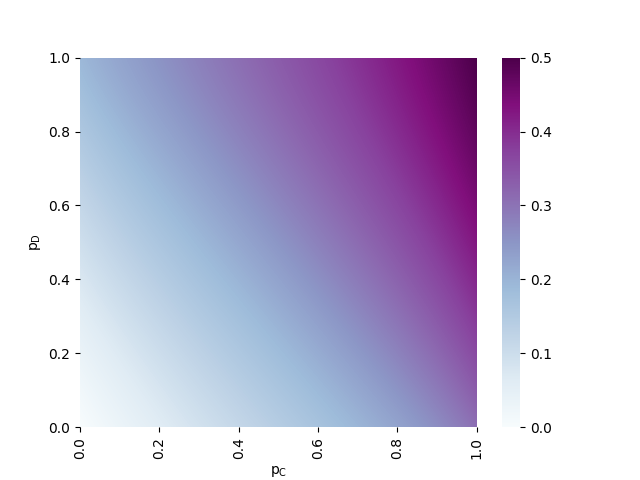}
         \caption{$\Mean{\statpr CC}$}
         \label{fig:piCC}
     \end{subfigure}
     \hfill
     \begin{subfigure}[b]{0.49\textwidth}
         \centering
         \includegraphics[width=\textwidth]{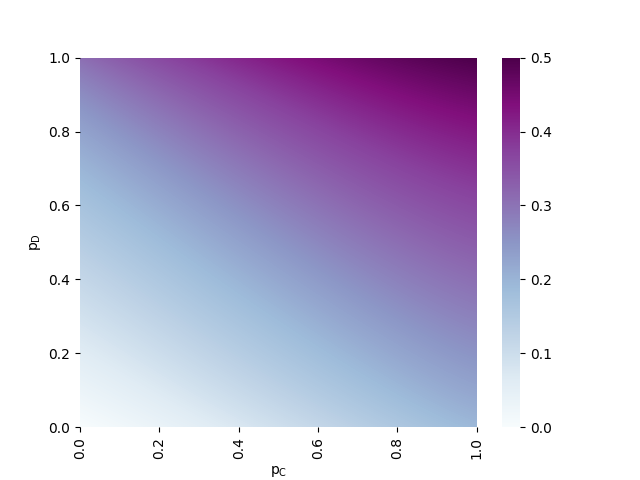}
         \caption{$\Mean{\statpr CD}$}
         \label{fig:piCD}
     \end{subfigure}
     \hfill
     \begin{subfigure}[b]{0.49\textwidth}
         \centering
         \includegraphics[width=\textwidth]{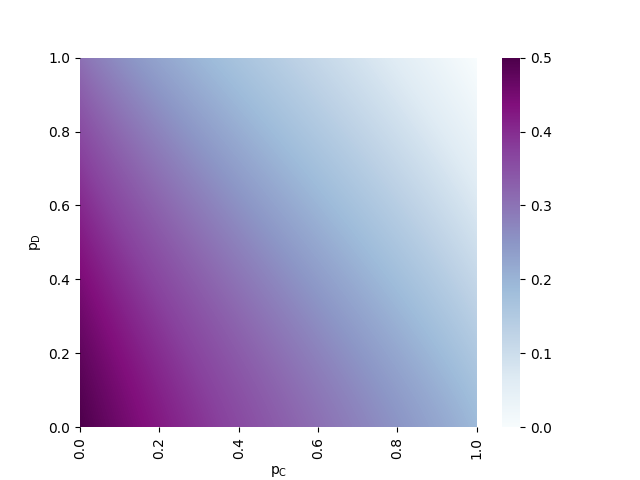}
         \caption{$\Mean{\statpr DC}$}
         \label{fig:piDC}
     \end{subfigure}
     \hfill
     \begin{subfigure}[b]{0.49\textwidth}
         \centering
         \includegraphics[width=\textwidth]{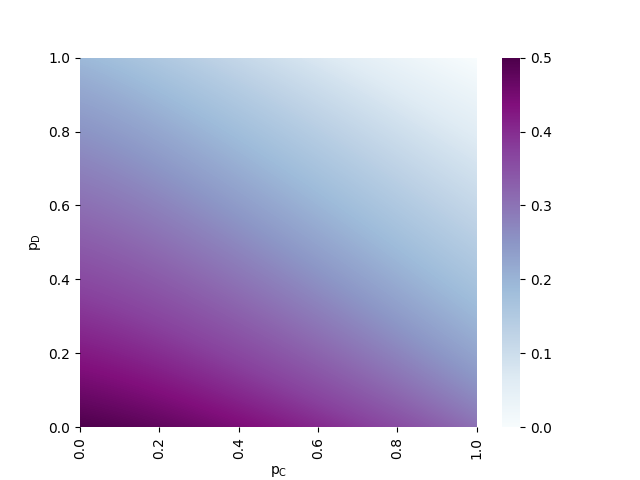}
         \caption{$\Mean{\statpr DD}$}
         \label{fig:piDD}
     \end{subfigure}
     \\
     \hfill
        \caption{Heat diagrams for $\Mean{c}$, $\Mean{c \prm}$, $\Mean{\statpr CC}$, $\Mean{\statpr CD}$, $\Mean{\statpr DC}$, and $\Mean{\statpr DD}$ (a,b,c,d,e,f) with respect to $\pr C$ and $\pr D$}
        \label{fig:heat maps 2}
\end{figure}

\begin{figure}
     \centering
     \begin{subfigure}[b]{0.49\textwidth}
         \centering
         \includegraphics[width=\textwidth]{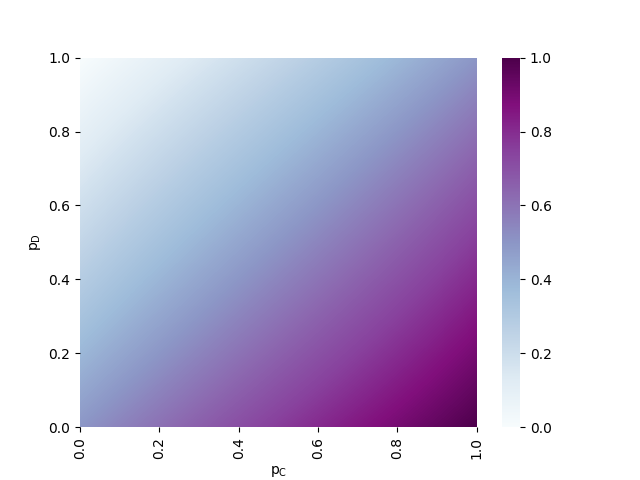}
         \caption{$\Mean{t}$}
         \label{fig:t}
     \end{subfigure}
     \hfill
     \begin{subfigure}[b]{0.49\textwidth}
         \centering
         \includegraphics[width=\textwidth]{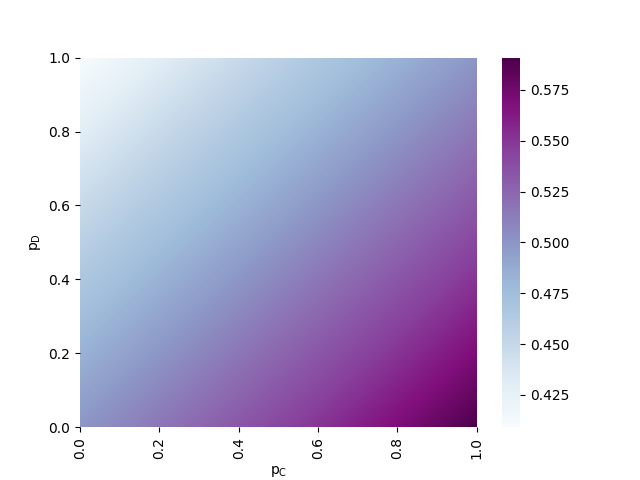}
         \caption{$\Mean{t \prm}$}
         \label{fig:t'}
     \end{subfigure}
     \hfill
     \begin{subfigure}[b]{0.49\textwidth}
         \centering
         \includegraphics[width=\textwidth]{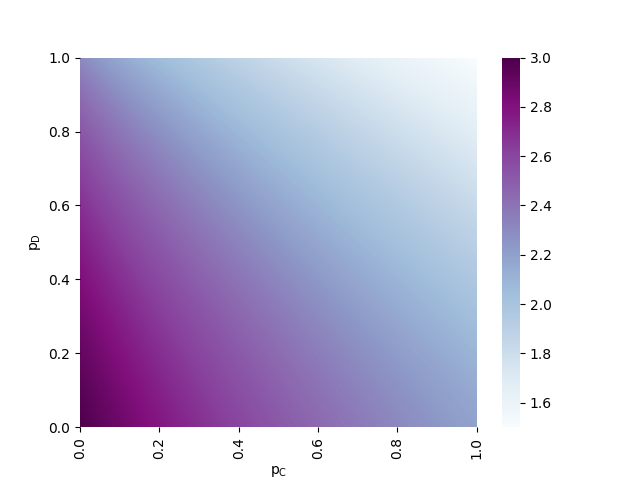}
         \caption{$\Mean{\score_{pr}}$}
         \label{fig:sPr}
     \end{subfigure}
     \hfill
     \begin{subfigure}[b]{0.49\textwidth}
         \centering
         \includegraphics[width=\textwidth]{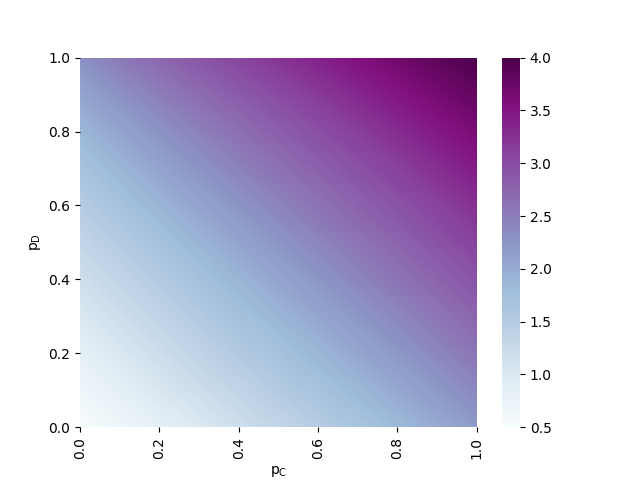}
         \caption{$\Mean{\scoreprm_{pr}}$}
         \label{fig:sPr'}
     \end{subfigure}
     \hfill
     \begin{subfigure}[b]{0.49\textwidth}
         \centering
         \includegraphics[width=\textwidth]{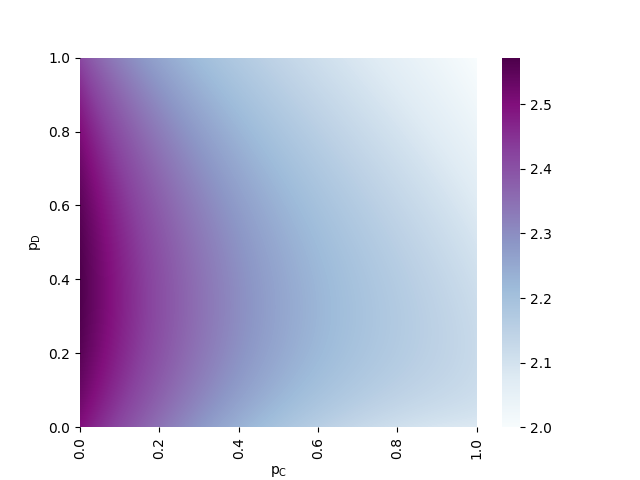}
         \caption{$\Mean{\score_{sn}}$}
         \label{fig:sSn}
     \end{subfigure}
     \hfill
     \begin{subfigure}[b]{0.49\textwidth}
         \centering
         \includegraphics[width=\textwidth]{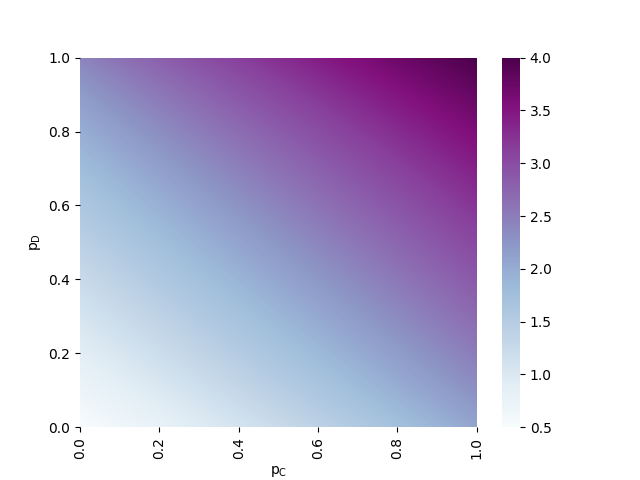}
         \caption{$\Mean{\scoreprm_{sn}}$}
         \label{fig:sSn'}
     \end{subfigure}
     \\
     \hfill
        \caption{Heat diagrams for $\Mean{t}$, $\Mean{t \prm}$, $\Mean{\score_{pr}}$, $\Mean{\scoreprm_{pr}}$, $\Mean{\score_{sn}}$, and $\Mean{\scoreprm_{sn}}$ (a,b,c,d,e,f) with respect to $\pr C$ and $\pr D$}
        \label{fig:heat maps 3}
\end{figure}

\begin{landscape}

These charts are intriguing, and already suffice to give us some useful information regarding the morality metrics.
\begin{enumerate}
\item In order to maximize one's score in the ISG, a player would always want to defect when their opponent cooperated in the last round. However, they would not want to defect every time their opponent defected. This suggests that that $\Mean{\score_{sn}}$ is more positively correlated with $\Mean{c}$ than $\Mean{\score_{pr}}$ is. 
\item Each of the reactive means, with the exception of scores, displays one or more symmetries. This is to be expected since there are no weights given to cooperation or defection until the scores are calculated. The symmetries evident in the stationary distributions are especially subtle since they do not display symmetry with respect to their own values, but rather with the plots of other stationary distributions.  
 
\item Asymptotic niceness, responsiveness, and reciprocity all have a positive correlation with $\pr C$, while a high score is anticorrelated with this value for both the IPD and ISG.

\item Both responsiveness and reciprocity increase as the chosen strategy becomes more like Tit-for-Tat. As we know from Axelrod's tournaments, this suggests that both of these metrics of justice are more likely to equate to victory over the opponent on average, as long as the ambient population is not excessively hostile.
\end{enumerate}

\subsection{Statistics}
We now compute means and standard deviations for our functions of interest as well as the naive values for the complements of interest. We compute these values exactly where convenient, and otherwise using 5,000 sample points distributed equally over the solution space. It can be seen that these values are concordant with the heatmaps above.
\begin{align*}
\begin{adjustbox}{max width = \columnwidth, center}
	\bgroup
	\def\arraystretch{1.3}
	\rowcolors{1}{lightgray}{}
	\begin{tabular}{ ccccccccccccccccc }
		$f$ & $\Mean{\pr C}$ & $\Mean{\pr D}$ & 
        $\Mean{\nice}$ &
        $\Mean{\coop}$ & $\Mean{\response}$ & $\Mean{\tft}$ & $\Mean{\scorepr}$ & $\Mean{\scoresn}$ & $\Mean{\statpr CC}$ & $\Mean{\statpr CD}$ & $\Mean{\statpr DC}$ & $\Mean{\statpr DD}$ \\
		$\mu(f)$ & $\frac 12$ & $\frac 12$ & $2-\frac{\pi^{2}}{6}$ & $\frac 12$ & $0$ & $\frac 12$ & $\frac 94$ & $\frac 94$ & $\frac 14$ & $\frac 14$ & $\frac 14$ & $\frac 14$ \\
		$\sigma(f)$ & $\frac{\sqrt{3}}{6}$ & $\frac{\sqrt 3}{6}$ & $0.256321$ & $0.206677$ & $0.408248$ & $0.206666$ & $0.312658$ & $0.127046$ & $0.105734$ & $0.105734$ & $0.105734$ & $0.105734$ 
  \\~\\
		$f$ & $\Mean{\niceprm}$ &  $\Mean{\coopprm}$ & $\Mean{\tftprm}$ & $\Mean{\scoreprprm}$ & $\Mean{\scoresnprm}$  \\
		$\mu(f)$ & $2-\frac{\pi^{2}}{6}$ & $\frac 12$ & $\frac 12$ & $\frac 94$ & $\frac 94$  \\
		$\sigma(f)$ & $0.118678$ & $0.009767$ & $0.0.03507$ & $0.723847$ & $0.726347$ 
	\end{tabular}
	\egroup
\end{adjustbox} 
\end{align*}
\\~\\
\newpage
Next, we compute covariances.
\begin{align*}
 \begin{adjustbox}{max width = \columnwidth, center}
		\bgroup
		\def\arraystretch{1.3}
		\rowcolors{1}{lightgray}{}		
		\begin{tabular}{ cccccccccc }
			$\text{cov}(\bullet, \bullet)$ & $\Mean{\pr C}$ & $\Mean{\pr D}$ & $\Mean{\nice}$ & $\Mean{\niceprm}$ & $\Mean{\coop}$ & $\Mean{\coopprm}$ & $\Mean{\tft}$ & $\Mean{\tftprm}$ & $\Mean{\statpr CC}$ \\
			$\Mean{\pr C}$ & $\frac 1{12}$ & $0$ & $\frac{-9+\pi^2}{12}$ & $-0.0326384$ & $\frac{-17 + 3 \pi^2 - 16 \log(2)}{36}$ & $0$ & $\frac{-17 + 3 \pi^2 - 16 \log(2)}{36}$ & $\frac{-38 + 5 \pi^2 - 16 \log(2)}{36}$ & $\frac{29 - 4 \pi^2 + 16 \log(2)}{24}$ \\
			$\Mean{\pr D}$ & $0$ & $\frac 1{12}$ & $0$ & $0$ & $\frac{-17 + 3 \pi^2 - 16 \log(2)}{36}$ & $0$ & $\frac{17 - 3 \pi^2 + 16 \log(2)}{36}$ & $\frac{38 - 5 \pi^2 + 16 \log(2)}{36}$ & $\frac{-121 + 18 \pi^2 - 80 \log(2)}{72}$ \\
			$\Mean{\nice}$ & $\frac{-9+\pi^2}{12}$ & $0$ & $0.0657006$ & $-0.0303034$ & $0.0367112$ & $0.000336208$ & $0.0367061$ & $0.00622641$ & $0.0223897$ \\
			$\Mean{\niceprm}$ & $-0.0326384$ & $0$ & $-0.0303034$ & $0.0140846$ & $-0.0165455$ & $-0.000222786$ & $-0.0165413$ & $-0.00280512$ & $-0.0101369$
            \\
            $\Mean{\coop}$ & $\frac{-17 + 3 \pi^2 - 16 \log(2)}{36}$ & $\frac{-17 + 3 \pi^2 - 16 \log(2)}{36}$ & $0.0367112$ & $-0.0165455$ & $0.0427154$ & $0$ & $0$ & $0$ & $\frac{25 \pi^2 + 24 \log (2) - 48 \log(2)^2 - 198 \zeta(3)}{108}$ \\
			$\Mean{\coopprm}$ & $0$ & $0$ & $0.000336208$ & $-0.000222786$ & $0$ & $\frac{849 - 98 \pi^2 + 864 \log(2) + 576 \log(2)^2 - 630 \zeta(3)}{1080}$ & $0$ & $0$ & $0.0000476943$ \\
			$\Mean{\tft}$ & $\frac{-17 + 3 \pi^2 - 16 \log(2)}{36}$ & $\frac{17 - 3 \pi^2 + 16 \log(2)}{36}$ & $0.0367061$ & $-0.0165413$ & $0$ & $0$ & $\frac{1849 + 532 \pi^2 + 5504 \log(2) + 1536 \log(2)^2 - 9630 \zeta(3)}{1800}$ & $\frac{953 + 239 \pi^2  + 448 \log(2) - 768 \log(2)^2 - 2700 \zeta(3)}{1080}$ & $0.00447272$ \\
			$\Mean{\tftprm}$ & $\frac{-38 + 5 \pi^2 - 16 \log(2)}{36}$ & $\frac{38 - 5 \pi^2 + 16 \log(2)}{36}$ & $0.00622641$ & $-0.00280512$ & $0$ & $0$ & $\frac{953 + 239 \pi^2  + 448 \log(2) - 768 \log(2)^2 - 2700 \zeta(3)}{1080}$ & $0.00122994$ & $0.000756751$ \\
            $\Mean{\statpr CC}$ & $\frac{29 - 4 \pi^2 + 16 \log(2)}{24}$ & $\frac{-121 + 18 \pi^2 - 80 \log(2)}{72}$ & $0.0223897$ & $-0.0101369$ & $\frac{25 \pi^2 + 24 \log (2) - 48 \log(2)^2 - 198 \zeta(3)}{108}$ & $0.0000476943$ & $0.00447272$ & $0.000756751$ & $0.0111797$ \\
		\end{tabular}
		\egroup
	\end{adjustbox}
 \end{align*}
 \\~\\

 We computed the mean and covariances of the auxiliary function $\coopprm$ in order to compute covariances for $\Mean{\score_{pr}}$, $\Mean{\score_{sn}}$, $\Mean{\statpr CD}$, $\Mean{\statpr DC}$, and $\Mean{\statpr DD}$. As covariance is bilinear, the listed covariances suffice to compute covariance of $\Mean{\response}$, $\Mean{\score_{pr}}$, $\Mean{\score_{sn}}$, $\Mean{\statpr CD}$, $\Mean{\statpr DC}$, and $\Mean{\statpr DD}$ with each of our functions of interest. We can now calculate the correlations.

\begin{align*}
    \begin{adjustbox}{max width = \columnwidth, center}
		\bgroup
		\def\arraystretch{1.5}
		\rowcolors{1}{lightgray}{}	
		\begin{tabular}{ cccccccccccccccccc }
			$\text{cor}(\bullet, \bullet)$ & $\Mean{\pr C}$ & $\Mean{\pr D}$ & $\Mean{\nice}$ &
            $\Mean{\niceprm}$& $\Mean{\coop}$ & $\Mean{\coopprm}$ &
            $\Mean{r}$ & $\Mean{\tft}$ & $\Mean{\tftprm}$ & $\Mean{\score_{\text{pr}}}$ & $\Mean{\scoreprm_{\text{pr}}}$ & $\Mean{\score_{\text{sn}}}$ & 
            $\Mean{\scoreprm_{\text{sn}}}$ & $\Mean{\statpr CC}$ & $\Mean{\statpr CD}$ & $\Mean{\statpr DC}$ & $\Mean{\statpr DD}$ 
            \\
			$\Mean{\pr C}$ & $1$ & $0$ & $0.979369$ & $-0.952681$ & $0.706966$ & $0$ & $0.707107$ & $0.707004$ & $0.706977$ & $-0.749826$ & $0.685406$ & $-0.935592$ & $0.641005$ & $0.835355$ & $0.546542$ & $-0.835355$ & $-0.546542$  \\
			$\Mean{\pr D}$ & $0$ & $1$ & $0$ & $0$ & $0.706966$ & $0$ &
            $-0.707107$ & $-0.707004$ & $-0.706977$ & $-0.652156$ & $0.727593$ & $-0.214496$ & $0.767132$ & $0.546542$ & $0.835355$ & $-0.546542$ & $-0.835355$   
            \\
            $\Mean{\nice}$ & $0.979369$ & $0$ & $1$ 
            & $-0.993105$ & $0.692982$ & $0.134300$ & $0.692532$ & $0.692923$ & $0.692647$ & $-0.720682$ & $0.668969$ & $-0.883691$ & $0.626942$ & $0.826131$ & $0.528430$ & $-0.813726$ & $-0.540835$
            \\
            $\Mean{\niceprm}$ & $-0.952681$ & $0$ & $-0.993105$ & $1$ & $-0.674552$ & $-0.192207$ & $-0.673647$ & $-0.674419$ & $-0.673965$ & $0.695069$ & $-0.649813$ & $0.845596$ & $-0.609493$ & $-0.807838$ & $-0.510712$ & $0.790067$ & $0.528456$  
            \\
			$\Mean{\coop}$ & $0.706966$ & $0.706966$ & $0.692982$ & $-0.674552$ & $1$ & $0$ &
            $0$ & $0$ & $0$ & $-0.991548$ & $0.999340$ & $-0.813397$ & $0.995900$ & $0.977342$ & $0.977342$ & $-0.977342$ & $-0.977342$   \\
			$\Mean{\coopprm}$ & $0$ & $0$ & $0.134300$
            & $-0.192207$ & $0$ & $1$ & $-0.000866735$ & $0$ & $0$ & $0.109332$ & $-0.0202392$ & $0.269065$ & $-0.00672317$ & $0.0461852$ & $-0.0461852$ & $0.0461852$ & $-0.0461852$  \\
            $\Mean{r}$ & $0.707107$ & $-0.707107$ & $0.692532$ & $-0.673647$& $0$ & $-0.000866735$ & $1$ & $0.999855$ & $0.999817$ & $-0.0691574$ & $-0.0298136$ & $-0.51005$ & $-0.0891795$ & $0.204183$ & $-0.204183$ & $-0.204259$ & $0.204259$
            \\
            $\Mean{\tft}$ & $0.707004$ & $0.707004$ & $0.692923$ & $-0.674419$ & $0$ & $0$ & $0.999855$ & $1$ & $0.999506$ & $-0.0692203$ & $-0.029899$ & $-0.511052$ & $-0.0893882$ & $0.204686$ & $-0.204686$ & $-0.204686$ & $0.204686$    \\
			$\Mean{\tftprm}$ & $0.706977$ & $-0.706977$ & $0.692647$ & $-0.673965$ & $0$ & $0$ & $0.999817$ & $0.999506$ & $1$ & $-0.0690147$ & $-0.0298102$ & $-0.509534$ & $-0.0891227$ & $0.204078$ & $-0.204078$ & $-0.204078$ & $0.204078$  \\
            $\Mean{\score_{\text{pr}}}$ & $-0.749826$ & $-0.652156$ & $-0.720682$ 
            & $0.695069$ & $-0.991548$ & $0.109332$ & $-0.0691574$ & $-0.0692203$ & $-0.0690147$ & $1$ & $-0.990998$ & $0.871966$ & $-0.981916$ & $-0.978462$ & $-0.959701$ & $0.988561$ & $0.949602$  \\
			$\Mean{\scoreprm_{\text{pr}}}$ & $0.685406$ & $0.727593$ & $0.668969$ & $-0.649813$ & $0.999340$ & $-0.0202392$ & $-0.0298136$ & $-0.029899$ & $-0.0298102$ & $-0.990998$ & $1$ & $-0.802744$ & $0.998101$ & $0.969530$ & $0.983865$ & $-0.971399$ & $-0.981995$   \\
			$\Mean{\score_{\text{sn}}}$ & $-0.935592$ & $-0.214496$ & $-0.883691$ & $0.845596$ & $-0.813397$ & $0.269065$ & $-0.51005$ & $-0.511052$ & $-0.509534$ & $0.871966$ & $-0.802744$ & $1$ & $-0.765347$ & $-0.889074$ & $-0.700860$ & $0.913928$ & $0.676006$   \\
			$\Mean{\scoreprm_{\text{sn}}}$ & $0.641005$ & $0.767132$ & $0.626942$ & $-0.609493$ & $0.995900$ & $-0.00672317$ & $-0.0891795$ & $-0.0893882$ & $-0.0891227$ & $-0.981916$ & $0.998101$ & $-0.765347$ & $1$ &$0.954391$ & $0.992280$ & $-0.955012$ & $-0.991659$   \\
			$\Mean{\statpr CC}$ & $0.835355$ & $0.546542$ & $0.826131$ & $-0.807838$ & $0.977342$ & $0.0461852$ & $0.204183$ & $0.204686$ & $0.204078$ & $-0.978462$ & $0.969530$ & $-0.889074$ & $0.954391$ & $1$ & $0.910396$ & $-0.995734$ & $-0.914662$   \\
			$\Mean{\statpr CD}$ & $0.546542$ & $0.835355$ & $0.528430$ & $-0.510712$ & $0.977342$ & $-0.0461852$ &$-204183$ & $-0.204686$ & $-0.204686$ & $-0.959701$ & $0.983865$ & $-0.700860$ & $0.992280$ & $0.910396$ & $1$ & $-0.914662$ & $-0.995734$   
            \\
			$\Mean{\statpr DC}$ & $-0.835355$ & $-0.546542$ & $-0.813726$ & $0.790067$ & $-0.977342$ & $0.0461852$ &
            $-0.0691574$ & $-0.204686$ & $-0.204078$ & $0.988561$ & $-0.971399$ & $0.913928$ & $-0.955012$ & $-0.995734$ & $-0.914662$ & $1$ & $0.910396$  \\
			$\Mean{\statpr DD}$ & $-0.546542$ & $-0.835355$ & $-0.540835$ & $0.528456$ & $-0.977342$ & $-0.0461852$ &
            $0.204259$ & $0.204686$ & $0.204078$ & $0.949602$ & $-0.981995$ & $0.676006$ & $-0.991659$ & $-0.914662$ & $-0.995734$ & $0.910396$ & $1$  \\
		\end{tabular}
		\egroup
	\end{adjustbox}
 \end{align*}
\end{landscape}

From the above chart, it is evident that cooperation rate and asymptotic niceness are strongly anticorrelated with success, especially the former. Niceness is slightly more beneficial in the IPD than in the ISG, while the opposite is true with respect to the cooperation rate. It is interesting that this type of behavior is observed in asymptotic niceness given that the results of Axelrod's tournaments suggest nice strategies have the unusually good performance.
\\
Another interesting observation is that the correlation between responsiveness and success in the IPD, while slightly negative, is almost zero. This same relationship can be seen with reciprocity, as $r$ and $\tft$ have virtually the same dynamics and correlations. This suggests that, if one views justice as treating others how they treat you, a player can play an almost perfectly just game and be victorious against around half of the reactive memory one strategies. This idea is additionally supported by the almost identical correlations for $t \prm$.

\section{Discussions and Conclusion}
\label{Section: Future Work} 
In this work we have focused on a few simple yet intuitive morality metrics, and it is straightforward to consider various extensions in this regard. Given the subjective nature of morality, there are numerous other functions to be investigated under the framework of this paper. For instance, if a player believes they are just when they treat their opponent as they are treated, a measure of morality could be the long-run probability of making the same move as their partner in the next round, or some other variation of this aspect. In other words, there could be a measure of the probability of a player's opponent and the player themselves cooperating in the same round or defecting in the same round. This idea is already partially captured by $\Mean{\statpr CC}$ and $\Mean{\statpr DD}$, so a linear combination of these two values could be a good metric for this idea.  
\\
Another potential function of interest would be a variation on positive reciprocity that just takes into account how often a player reciprocates when their opponent cooperates. Additionally, the different functions could be broken down into different distributions to see exactly how moral one needs to be to succeed. For instance, one could examine $\response \geq 0$ against $\response \leq 0$. This would specifically convey whether or not it is better to cooperate more with cooperators than defectors or vice versa. This idea would be especially useful for $r$ and $t$ since their correlations with score were so small in absolute value.
\\
Beyond this, one could examine an environment where an opponent or a player is more likely to choose one strategy over another. In other words, instead of assuming the opponent chooses a strategy with uniform probability, the distribution could be a truncated Gaussian or another distribution. Similarly, a fine-grained description of their pairwise encounters (who-meets-whom relationships) can be based on graphs or networks~\cite{fu2009evolutionary}. This incorporation could help reflect the tendency of certain populations to congregate when they have shared ideals.
\\
While the focus of this paper was on morality, the development of reactive means has more widespread applications. Any function that measures the behavior between two competing players can theoretically have its reactive mean computed. This could lead to analyses on other ideas such as the consistency of a player's moves~\cite{van2015importance} or the success of a player past just their average score.
\\

Lastly, we could take averages over families of opponents besides reactive memory one strategies or players having asymmetrical roles~\cite{izgi2023extended}. It would be natural, for instance, to integrate against all memory one strategies: arguably, this level of generality suffices for a total understanding of the IPD \cite{press2012iterated}. 

If desired, one could of course integrate over finite-memory strategies~\cite{hauert1997effects}, or any family of strategies which bears a natural parameterization. In an adaptation of the EigenJesus and EigenMoses metrics derived in \cite{Singer}, one could determine the scores for a player in an environment of opponents sampled independently from the pool of all reactive memory one strategies. Under this framework, as the number of opponents increases, the measured scores will approach their reactive average values. In addition, the development of simple and intuitive metrics using reactive means will aid in evaluating and comparing IPD strategies generated in complicated ways, such as those based on machine learning algorithms, including reinforcement learning~\cite{harper2017reinforcement,sandholm1996multiagent} and particle swarm optimization~\cite{franken2005particle}. 

In sum, we evaluate and compare the moral nature of reactive strategies employed in the Iterated Prisoner's Dilemma (IPD) by drawing on  human's intuitive perception of ``fair play'' and ``goodness''. Using these morality metrics, we demonstrate that two of the metrics are significantly associated with their success in the IPD, while the other two metrics are weakly related. Our results can help further conceive new ways, by means of integrating morality concerns, for enhancing fairness and cooperation among adaptive and learning individuals~\cite{barfuss2020caring,mcavoy2022selfish}.

\section*{Code Availability}
The code used in this study is available upon reasonable request.

\section*{Acknowledgments}
We thank Steve Fan for his clever comments on integration. F.F. gratefully acknowledges support from the Bill \& Melinda Gates Foundation (award no. OPP1217336), the NIH COBRE Program (grant no.1P20GM130454), and the Neukom CompX Faculty Grant.

\bibliography{refs} 

\newpage
\section*{Appendix}
\label{Section: Appendix}
The reactive means of the functions we defined in Section \ref{Section: Metrics of Justice} may be computed numerically for fixed $\pr C$ and $\pr D$, but these computations become much slower as we permit $\pr C$ and $\pr D$ to vary. Moreover, a double integral with both $\pr C$ and $\pr D$ in the denominator is difficult (but possible) to integrate symbolically.

However, if $\pr C \neq \pr D$, we may write $\responseprm = \prprm C - \prprm D$ and perform the change of variables $\prprm C = \responseprm + \prprm D$ to obtain 
\begin{align}
	\Mean f(A) &= \int\limits_{-1}^0 \int\limits_{-\response\prm}^{1} f\parent{\pr C, \pr D, \responseprm + \prprm D, \prprm D} \dif {\prprm D} \dif {\responseprm} \nonumber \\
	&+ \int\limits_0^1 \int\limits_{0}^{1-\response\prm} f\parent{\pr C, \pr D, \responseprm + \prprm D, \prprm D} \dif {\prprm D} \dif {\responseprm}. \nonumber 
\end{align}

This change of variables renders our integrals much more manageable, and a straightforward but tedious computation yields

\begin{align*}
    \Mean \nice (A) &= 1 + \frac{\parent{1 - \pr C}}{\pr C} \log\parent{1 - \pr C} \\
	\Mean {\coop} (A) &= \frac{1}{2} \\
	&+ \frac{(1 + \response) (-1 + 2 \pr D + \response) \log\parent{1 + \response}}{2 \response^2} \\
	&+ \frac{(1 - \response) (-1 + 2 \pr D + \response) \log\parent{1 - \response}}{2 \response^2} \\
	\Mean {\coopprm} (A) &= \frac{1 - 2 \pr D}{2 \response} \\
	&+ \frac{(1 + \response) (-1 + 2 \pr D + \response) \log\parent{1 + \response}}{2 \response^3} \\
	&+ \frac{(1 - \response) (-1 + 2 \pr D + \response) \log\parent{1 - \response}}{2 \response^3} \\
	\Mean \tft (A) &= \frac{3}{2} + \frac{-1 + 4 \pr D - 4 \pr D^2 - 4 \pr D \response}{2 \response} \\
	&+ \frac{(1 + \response) (-1 + 2 \pr D + \response)^2 \log\parent{1 + \response}}{2 \response^3} \\
	&+ \frac{(1 - \response) (-1 + 2 \pr D + \response)^2 \log\parent{1 - \response}}{2 \response^3} \\ 
 	\Mean {\statpr CC} (A) &= \frac{2 - \pr D}{2 \response} \\
 	&+ \frac{\left[ \parent{-2 + 5 \pr D + 2 \response - 4 \pr D^2 - 3 \pr D \response - \response^2} + \response \parent{-2 + 3 \pr D + \response - 2 \pr D^2 - \pr D \response}\right]\log\parent{1 + \response}}{2 \response^3} \\
 	&+ \frac{\left[ \parent{-2 + 5 \pr D + 2 \response - 4 \pr D^2 - 3 \pr D \response - \response^2} - \response \parent{-2 + 3 \pr D + \response - 2 \pr D^2 - \pr D \response} \right] \log\parent{1 - \response}}{2 \response^3} \\
 \end{align*}
 \begin{align*}
	\Mean {\statpr CD} (A) &= \frac 12 - \frac{2 - \pr D}{2 \response} \\
 	&+ \frac{\left[ \parent{2 - 5 \pr D - 3 \response + 4 \pr D^2 + 5 \pr D \response + 2 \response^2} + \response \parent{2 - 3 \pr D - 2 \response + 2 \pr D^2 + 3 \pr D \response + \response^2} \right] \log\parent{1 + \response}}{2 \response^3} \\
	&+ \frac{\left[ \parent{2 - 5 \pr D - 3 \response + 4 \pr D^2 + 5 \pr D \response + 2 \response^2} - \response \parent{2 - 3 \pr D - 2 \response + 2 \pr D^2 + 3 \pr D \response + \response^2} \right] \log\parent{1 - \response}}{2 \response^3} \\
	\Mean {\statpr DC} (A) &= -\frac{1 + \pr D}{2 \response} \\
 	&+ \frac{\left[ \parent{1 - 3 \pr D - \response + 4 \pr D^2 + 3 \pr D \response + \response^2} + \response \parent{1 - \pr D + 2 \pr D^2 + \pr D \response} \right] \log\parent{1 + \response}}{2 \response^3} \\
	&+ \frac{\left[ \parent{1 - 3 \pr D - \response + 4 \pr D^2 + 3 \pr D \response + \response^2} - \response \parent{1 - \pr D + 2 \pr D^2 + \pr D \response} \right] \log\parent{1 - \response}}{2 \response^3} \\
	\Mean {\statpr DD} (A) &= \frac{1}{2} + \frac{1 + \pr D}{2 \response} \\
 	&+ \frac{\left[\parent{-1 + 3 \pr D + 2 \response - 4 \pr D^2 - 5 \pr D \response - 2 \response^2} + \response \parent{-1 + \pr D + \response - 2 \pr D^2 - 3 \pr D \response - \response^2} \right] \log\parent{1 + \response}}{2 \response^3} \\
	&+ \frac{\left[\parent{-1 + 3 \pr D + 2 \response - 4 \pr D^2 - 5 \pr D \response - 2 \response^2} - \response \parent{-1 + \pr D + \response - 2 \pr D^2 - 3 \pr D \response - \response^2} \right] \log\parent{1 - \response}}{2 \response^3} \\
\end{align*}

Here we have adopted that $\log x$ refers to the natural logarithm, rather than $\log_2 x$ or $\log_{10} x$.

Of course $\Mean {\pr C} = \pr C$, $\Mean{\pr D} = \pr D$, and $\Mean{\response} = \response$. The functions $\Mean{\scorepr}$ and $\Mean{\scoresn}$ may be computed as linear combinations of $\Mean{\statpr CC}$, $\Mean{\statpr CD}$, $\Mean{\statpr DC}$, and $\Mean{\statpr DD}$. These formulas are valid except when $\response \in \set{-1, 0, 1}$. But $\response = 0$ if and only if $\pr C = \pr D$, and in this case \eqref{fA defined independent of p0} suffices to compute the reactive mean of each function in our set above. Indeed, the denominators in \eqref{c defined} and \eqref{cprm defined} simplify to 1, and we have

\begin{center}
    \begin{tabularx}{\linewidth}{@{} lCCr @{}}
\refstepcounter{equation}
    &   $\begin{aligned}
    \Mean \nice (A) &= 1 + \frac{\parent{1 - \pr C}}{\pr C} \log\parent{1 - \pr C} \\
	\Mean {\coop} (A) &= \pr C \\
	\Mean {\coopprm} (A) &= \frac{1}{2} \\
	\Mean \tft (A) &= \frac 12
        \end{aligned}$
        &   $\begin{aligned}
 	\Mean {\statpr CC} (A) &= \frac{\pr C}{2} \\
	\Mean {\statpr CD} (A) &= \frac{\pr C}{2} \\
	\Mean {\statpr DC} (A) &= \frac{1 - \pr C}2 \\
	\Mean {\statpr DD} (A) &= \frac{1 - \pr C}2
            \end{aligned}$
            &   \refstepcounter{equation}
    \end{tabularx}
    \end{center}

Now if $\response \pm 1$, then $A$ is Tit-for-Tat ($\pr C = 1$, $\pr D = 0$) or the Bully $(\pr C = 0$, $\pr D = 1$), respectively. But these boundary cases are easy to evaluate directly. Recalling that $T$ denotes Tit-for-Tat, and $B$ denotes the Bully, we have

\begin{center}
    \begin{tabularx}{\linewidth}{@{} lCCr @{}}
\refstepcounter{equation}
    &   $\begin{aligned}
    \Mean \nice (T) &= 1 \\
	\Mean {\coop} (T) &= \frac 12 \\
	\Mean {\coopprm} (T) &= \frac 12 \\
	\Mean \tft (T) &= 1 \\
 	\Mean {\statpr CC} (T) &= 1 - \log(2) \\
	\Mean {\statpr CD} (T) &= - \frac 12 + \log(2)  \\
	\Mean {\statpr DC} (T) &= - \frac 12 + \log(2) \\
	\Mean {\statpr DD} (T) &= 1 - \log(2)
        \end{aligned}$
        &   $\begin{aligned}
    \Mean \nice (B) &= 0 \\
	\Mean {\coop} (B) &= \frac 12 \\
	\Mean {\coopprm} (B) &= \frac 12 \\
	\Mean \tft (B) &= 0 \\
 	\Mean {\statpr CC} (B) &= - \frac 12 + \log(2) \\
	\Mean {\statpr CD} (B) &= 1 - \log(2) \\
	\Mean {\statpr DC} (B) &= 1 - \log(2) \\
	\Mean {\statpr DD} (B) &= - \frac 12 + \log(2)
            \end{aligned}$
            &   \refstepcounter{equation}
    \end{tabularx}
    \end{center}

Irrespective of these considerations, we inherit the relations from Sections \ref{Section: Introduction} and \ref{Section: Metrics of Justice}, that is, from the behavior of the functions themselves.
\begin{align*}
	\Mean {\pr C} (A) &= \pr C \\
	\Mean {\pr D} (A) &= \pr D \\
	\Mean {\response} (A) &= \response \\
    \Mean{\tft} (A) &= \pr C \cdot \Mean{\coopprm}(A) + (1 - \pr D) \cdot \parent{1 - \Mean{\coopprm}(A)} \\
    \Mean \score (A) &= R_{CC} \cdot \Mean{\statpr CC} (A) + R_{CD} \cdot \Mean{\statpr CD} (A) + R_{DC} \cdot \Mean{\statpr DC} (A) + R_{DD} \cdot \Mean{\statpr DD} (A) \\
	\Mean {\statpr CD} (A) &= \Mean{\coop} (A) - \Mean{\statpr CC}(A) \\
	\Mean {\statpr DC} (A) &= \Mean{\coopprm} (A) - \Mean{\statpr CC}(A) \\
	\Mean {\statpr DD} (A) &= 1 - \Mean{\coop} (A) - \Mean{\coopprm} (A) + \Mean{\statpr CC}(A) \\
\end{align*}
With these identities in hand, it is now straightforward to generate the heatmaps that will be given in Section \ref{Section: Results and Analysis} (the authors used Python). These expressions also suffice to compute exact values for most of covariances between our functions of interest. As an example, we compute the covariance of $\Mean {\pr C}$ and $\Mean {\coop}$. Explicitly, we have 

\begin{align}
	\text{cov}\parent{\Mean{\pr C}, \Mean{\coop}} &= \frac{1}{\vol(X)} \int\limits_X \parent{\Mean{\pr C}(A) - \frac 12} \parent{\Mean{\coop} (A) - \frac 12} \dif A \nonumber \\
	&= \int\limits_{-1}^0 \int\limits_{-\response}^{1} \parent{\Mean{\pr C}\parent{\response + \pr D, \pr D} - \frac 12} \parent{\Mean{\coop}\parent{\response + \pr D, \pr D} - \frac 12} \dif {\pr D} \dif {\response} \nonumber \\
	&+ \int\limits_{-1}^0 \int\limits_{-\response}^{1} \parent{\Mean{\pr C}\parent{\response + \pr D, \pr D} - \frac 12} \parent{\Mean{\coop}\parent{\response + \pr D, \pr D} - \frac 12} \dif {\pr D} \dif {\response} \nonumber \\
	&= \int\limits_{-1}^0 \frac{\parent{1 + \response}^4 \log\parent{1 + \response} + \parent{1 + \response}^3 \parent{1 - \response} \log\parent{1 - \response}}{12 \response^2} \dif \response \nonumber \\
	&+ \int\limits_0^1 \frac{\parent{1 - \response}^4 \log\parent{1 - \response} + \parent{1 - \response}^3 \parent{1 + \response} \log\parent{1 + \response}}{12 \response^2} \dif \response \nonumber \\
	&= 2 \int\limits_0^1 \frac{\parent{1 - \response}^4 \log\parent{1 - \response} + \parent{1 - \response}^3 \parent{1 + \response} \log\parent{1 + \response}}{12 \response^2} \dif \response. \nonumber
	\intertext{For ease of exposition, write $\phi(\response)$ for the integrand of this last integral. We have}
	\phi(\response) &= \frac{\parent{1 - \response}^4 \log\parent{1 - \response} + \parent{1 - \response}^3 \parent{1 + \response} \log\parent{1 + \response}}{12 \response^2} \nonumber \\
	&= - \frac{\parent{1 - \response}^4}{12 \response^2} \sum\limits_{j = 1}^\infty \frac{\response^j}{j} + \frac{\parent{1 - \response}^3 \parent{1 + \response}}{12 \response^2} \sum\limits_{j = 1}^\infty \frac{(-1)^{j+1} \response^j}{j} \nonumber \\
	&= \frac{-1 + 4 \response - 6 \response^2 + 4 \response^3 - \response^4}{12 \response^2} \sum\limits_{j = 1}^\infty \frac{\response^j}{j} + \frac{-1 + 2 \response - 2 \response^3 + \response^4}{12 \response^2} \sum\limits_{j = 1}^\infty \frac{(-1)^{j} \response^j}{j} \nonumber \\
	&= \sum\limits_{j = 1}^\infty \parent{\frac{-1 - (-1)^{j}}{12 j}} \response^{j-2} + \sum\limits_{j = 1}^\infty \parent{\frac{2 + (-1)^j }{6 j}} \response^{j-1} - \sum\limits_{j = 1}^\infty \parent{\frac{1}{2 j}} \response^{j} \nonumber \\
	&+ \sum\limits_{j = 1}^\infty \parent{\frac{2-(-1)^j}{6 j}} \response^{j+1} + \sum\limits_{j = 1}^\infty \parent{\frac{-1 + (-1)^j}{12 j}} \response^{j+2} \nonumber 
	\intertext{Note that the coefficient of $\response\inv$ in the first sum vanishes, so $\phi(\response)$ is holomorphic at $0$. Now integrating each sum termwise, reindexing, and combining like terms, $\int\limits_0^1 \phi(\response)$ becomes}
	&- \sum\limits_{j = 1}^\infty \frac{7}{12 j (j+1)} + \sum\limits_{j = 1}^\infty \frac{1}{3 j(j+2)} - \sum\limits_{j = 1}^\infty \frac{1}{12 j(j+3)} \label{Telescoping terms} \\
	&+ \sum\limits_{j = 1}^\infty \frac{(-1)^{j}}{12 j (j+1)} - \sum\limits_{j = 1}^\infty \frac{(-1)^j}{6 j(j+2)}  + \sum\limits_{j = 1}^\infty \frac{(-1)^j}{12 j(j+3)} \label{Canceling log terms} \\
	&+ \sum\limits_{j = 1}^\infty \frac{1}{3 j^2} + \sum\limits_{j = 1}^\infty \frac{(-1)^j }{6 j^2} \label{pi^2 terms}
\end{align}
The series in \eqref{Telescoping terms} telescope to $- 7/12$, $1/4$, and $- 11/216$ respectively. Similarly, performing a partial fraction decomposition on the terms in the series of \eqref{Canceling log terms} yields scaled copies of the sum 
\[
\sum\limits_{j=1}^\infty \frac{(-1)^{j}}{j} = -\log \parent 2,
\] 
possibly with some terms omitted at the beginning of the series. This perspective lets us evaluate these series as $\frac 1{12} - \frac{1}{6} \log(2)$, $\frac{1}{24}$, and $\frac{5}{216} - \frac{1}{18} \log(2)$. Finally, classic methods from analytic number theory \cite{Apostol} let us evaluate the series in \eqref{pi^2 terms} as $\frac{\pi^2}{18}$ and $- \frac{\pi^2}{72}$. Summing these values, we conclude that
\[
	\int\limits_0^1 \phi(\response) \dif \response = \frac{-17 + 3 \pi^2 - 16 \log(2)}{72},
\]
and so
\[
	\text{cov}\parent{\Mean{\pr C}, \Mean{\coop}} = \frac{-17 + 3 \pi^2 - 16 \log(2)}{36}.
\]
The other covariances we give explicit values for can be evaluated similarly.

\end{document}

%% file: Alphabet_Commands.tex
\newcommand{\bbZ}{\mathbb{Z}}

\newcommand{\calF}{\mathcal{F}}

%% file: Math_Commands.tex
\newcommand{\set}[1]{\left\{{#1}\right\}}
\newcommand{\prm}{^\prime}
\newcommand{\parent}[1]{\left( #1 \right)}

\newcommand{\inv}{^{-1}}

\newcommand{\Zmod}[1]{\faktor{\bbZ}{n \bbZ}}

\newcommand{\vol}{{\text {\rm vol}}}

%% file: Paper_Specific_Commands.tex
\newcommand{\dif}{\text{\rm d}}

\newcommand{\score}{s} 
\newcommand{\scorepr}{\score_{\text{pr}}} 
\newcommand{\scoresn}{\score_{\text{sn}}} 
\newcommand{\scoreprprm}{\score_{\text{pr}\prm}} 
\newcommand{\scoresnprm}{\score_{\text{sn}\prm}} 

\newcommand{\pr}[1]{p_{#1}} 

\newcommand{\statpr}[2]{\pi_{#1 #2}} 
\newcommand{\coop}{c} 
\newcommand{\tft}{t} 
\newcommand{\nice}{n} 

\newcommand{\response}{r}
\newcommand{\responseprm}{r\prm}

\newcommand{\scoreprm}{s^{\prime}} 

\newcommand{\Mean}[1]{\overline{#1}}

\newcommand{\prprm}[1]{p^{\prime}_{#1}} 

\newcommand{\statprprm}[2]{\pi\prm_{#1 #2}} 
\newcommand{\coopprm}{c\prm} 
\newcommand{\tftprm}{t\prm} 
\newcommand{\niceprm}{n\prm} 